\documentclass[12pt,thmsa]{article}
\usepackage{amssymb}

\usepackage{sw20lart}

\input{tcilatex}
\begin{document}

\title{Quantum vacuum effects as generalized f(R) gravity. Application to stars.}
\author{Emilio Santos \\
Departamento de F\'{i}sica. Universidad de Cantabria. Santander. Spain.}
\maketitle

\begin{abstract}
It is assumed that, for weak spacetime curvature, the main gravitational
effect of the quantum vacuum stress-energy corresponds to adding two terms
to the Einstein-Hilbert action, proportional to the square of the curvature
scalar and to the contraction of two Ricci tensors, respectively. It is
shown that compatibility with terrestrial and solar systems observations
implies that the square roots of the coefficients of these terms should be
either a few millimeters or a few hundred meters. It is shown that the
vacuum contribution increase the stability of massive white dwarfs.

\end{abstract}

\section{ Introduction}

In the study of quantum fields in curved spacetime it has been stablished
that the quantum vacuum gives rise to a finite, non-zero, energy\cite
{Birrell}, \cite{Wald}. Furthermore, some effects have been attributed to
the gravity of the quantum vacuum, like the observed acceleration in the
expansion of the universe\cite{Sahni}.

The gravitational effects of the quantum vacuum may be taken into account
introducing a vacuum stress-energy tensor, $T_{\mu \nu }^{vac},$ in the
Einstein equation, which should read 
\begin{equation}
R_{\mu \nu }-\frac{1}{2}g_{\mu \nu }R=k\left( T_{\mu \nu }+T_{\mu \nu
}^{vac}\right) ,  \label{E}
\end{equation}
where $R_{\mu \nu }$ is the Ricci tensor, $R$ the curvature scalar and $%
T_{\mu \nu }$ is the stress-energy tensor of matter either baryonic or dark
plus radiation, $k$ is $8\pi $ times Newton\'{}s constant and we shall use
units $c=1$ throughout. We assume that the vacuum tensor, $T_{\mu \nu
}^{vac},$ depends on the space-time curvature, therefore it should be a
functional of the Riemann tensor, $R_{\mu \nu \lambda \sigma }$ (and the
metric, $g_{\mu \nu }.)$ It is plausible to derive the functional from a
generalized Einstein-Hilbert action 
\begin{equation}
S=\frac{1}{2k}\int d^{4}x\sqrt{-g}\left( R+F\right) +S_{mat},  \label{0a}
\end{equation}
where $F$ is associated to the vacuum. In general $F$ should be a function
of the scalars which may be obtained by combining the Riemann tensor, $%
R_{\mu \nu \lambda \sigma },$ and its derivatives, with the metric tensor, $%
g_{\mu \nu }$. Here I shall considers scalars more general than the Ricci
scalar, whose functions have been extensively explored in recent years under
the name of \textit{f(R)-gravity}\cite{Faraoni},\cite{Odintsov}.

The action eq.$\left( \ref{0a}\right) $ may be interpreted as a modification
of general relativity (RG), that is we may put the tensor $T_{\mu \nu
}^{vac} $ on the left side, rather than the right side, of Einstein eq.$%
\left( \ref{E}\right) $. Indeed it is equivalent in practice whether we
assume that the quantum vacuum gives rise to some stress-energy to be added
to the matter one or we assume that GR should be modified by adding to $R$ a
term $F$ in the gravitational action. Even if we remain within metric
theories of gravity, that is we assume that the stress-energy tensor of
matter produces the curvature of space-time, the arguments leading to the GR
choice (the standard Einstein equation) are not compulsory, but small
modifications are compatible with observations. That is we might introduce
the sum $R+F$, instead of the Ricci scalar $R$ in the Einstein-Hilbert
action. Of course there are strong constraints to the form of $F$ from both
observational evidence and requirements of consistence.

There are a number of proposals for the gravity of the quantum vacuum
derived from fundamental arguments involving the quantization of model
vacuum fields\cite{Parker}. They lead to $F$ being a function of scalars
like $R_{\mu \nu }R^{\mu \nu },R_{\mu \nu \lambda \sigma }R^{\mu \nu \lambda
\sigma }$ and $\Box R$ in addition to $R$. Most of the proposals have been
made for the study of cosmology, in particular the attempt to explain the
observed accelerated expansion of the universe (the ``dark energy''),
although it is not necessary to modify GR in order to explain it\cite{Santos}%
.

The aim of the present paper is to study the influence of modified GR in the
structure of Newtonian or weakly relativistic stars. A study of fully
relativistic stars has been made recently within f(R)-gravity\cite{Maeda},
where the authors conclude that neutron stars are not possible (or they
require extreme fine tuning) within \textit{f(R)} theories if these are
compatible with observational constraints. However the theories studied here
are more general than those cosidered in Ref.\cite{Maeda}.

\section{Vacuum gravity for weak curvature}

In order to find the most appropriate function $F$ to be put in the action
eq.$\left( \ref{0a}\right) $ I shall not attempt to derive it from
fundamental arguments, but use a plausible phenomenological approach
combining arguments of simplicity with dimensional considerations. The Ricci
scalar has dimensions, $L^{-2}$, of inverse squared length and the theory
derived by using it in the action (that is GR) is known to give very good
agreement with observations for a wide range of intensities of the
gravitational field (i.e. curvature of space-time.) The purpose of this
paper is to study the influence of vacuum gravity on Newtonian or weakly
relativistic stars. Therefore I propose to include in $F$ only terms with
dimensions not departing too much from $L^{-2}.$ Thus I shall assume that $F$
is a sum of terms with dimensions $L^{0},$ $L^{-2}$ and $L^{-4}.$ For
simplicity I will exclude more complicated terms with dimensions
intermediate between $L^{0}$ and $L^{-4}$, like $L^{-2}\log L$. Thus I am
lead to the following 
\begin{equation}
F=\Lambda +a_{0}R+a_{1}R^{2}+a_{2}R_{\mu \nu }R^{\mu \nu }+a_{3}\Box
R+a_{4}\nabla ^{\mu }\nabla ^{\nu }R_{\mu \nu }+a_{5}R_{\mu \nu \lambda
\sigma }R^{\mu \nu \lambda \sigma },  \label{f1}
\end{equation}
where $\nabla _{\mu }$ means the covariant derivative and $\Box $ $\equiv
g^{\mu \nu }\nabla _{\mu }\nabla _{\nu }$. The constant parameter $\Lambda $
has dimensions $L^{-2},$ $a_{0}$ is dimensionless and all remaining
coefficients $a_{j}$ have dimensions $L^{2}$. There are no other scalar
terms with similar dimensional dependence so that eq.$\left( \ref{f1}\right) 
$ gives the most general $F$ fulfilling the requirements of dimensionality
and simplicity above stated. However I recognize that the stated criterion
of simplicity might be questioned.

In eq.$\left( \ref{f1}\right) $ the term $\Lambda $ will give rise to a
``cosmological constant'' in Einstein\'{}s equation. It may be relevant in
cosmology, but it will have a negligible effect in the structure of stars
and I ignore it in the following. The term $a_{0}R$ may be absorbed in the
standard GR term, which amounts to a rescaling of Newton constant (the new
constant $k$ will be the old one divided by $1+a_{0}$). After this rescaling
the gravitational action in eq.$\left( \ref{0}\right) $ contains only the
``true vacuum polarization'', in the words of Zeldovich\cite{Zel}.. The term
with $a_{3}$ may be removed because it gives no contribution to the field
equations. The same is true for the term with $a_{4}$ due to the fact that $%
\nabla ^{\nu }(R_{\mu \nu }-\frac{1}{2}g_{\mu \nu }R)=0,$ which leads to the
equality $\nabla ^{\mu }\nabla ^{\nu }R_{\mu \nu }=\frac{1}{2}\Box R.$
Finally the term with $a_{5}$ may be removed taking into account the well
known fact that the Gauss-Bonnet invariant, 
\[
\mathcal{G}=R^{2}-4R_{\mu \nu }R^{\mu \nu }+R_{\mu \nu \lambda \sigma
}R^{\mu \nu \lambda \sigma }, 
\]
does not contribute to the field equations and therefore we may substitute $%
4R_{\mu \nu }R^{\mu \nu }-R^{2}$ for $R_{\mu \nu \lambda \sigma }R^{\mu \nu
\lambda \sigma }$ without changing those equations. Thus I will study here
the action

\begin{equation}
S=\frac{1}{2k}\int d^{4}x\sqrt{-g}\left( R+aR^{2}+bR_{\mu \nu }R^{\mu \nu
}\right) +S_{mat},  \label{0}
\end{equation}
which contains just two free parameters, $a$ and $b$.

The field equations associated to the action eq$.\left( \ref{0}\right) $ may
be got from Carroll et al.\cite{Turner} leading to 
\begin{eqnarray}
&&R_{\mu \nu }-\frac{1}{2}g_{\mu \nu }R+2a\left[ RR_{\mu \nu }-\frac{1}{4}%
g_{\mu \nu }R^{2}-\nabla _{\mu }\nabla _{\nu }R+g_{\mu \nu }\Box R\right] 
\nonumber \\
&&+b\left[ -\frac{1}{2}g_{\mu \nu }R_{\lambda \sigma }R^{\lambda \sigma
}+2R_{\mu }^{\sigma }R_{\sigma \nu }+\Box R_{\mu \nu }+g_{\mu \nu }\nabla
_{\lambda }\nabla _{\sigma }R^{\lambda \sigma }\right]  \nonumber \\
&&-b\left[ \nabla _{\sigma }\nabla _{\nu }R_{\mu }^{\sigma }+\nabla _{\sigma
}\nabla _{\mu }R_{\nu }^{\sigma }\right] =kT_{\mu \nu },  \label{1}
\end{eqnarray}
where $T_{\mu \nu }$ is the stress-energy tensor of matter. The last term of
the left side may be transformed taking into account the standard rule for
the commutation of covariant derivatives of a tensor, that is 
\begin{eqnarray}
\nabla _{\sigma }\nabla _{\nu }R_{\mu }^{\sigma } &=&g^{\lambda \sigma
}\nabla _{\sigma }\nabla _{\nu }R_{\mu \lambda }=\nabla _{\nu }\nabla
_{\sigma }R_{\mu }^{\sigma }+R_{\mu }^{\lambda }R_{\lambda \nu \sigma
}^{\sigma }-R_{\lambda }^{\sigma }R_{\mu \nu \sigma }^{\lambda }  \nonumber
\\
&=&R_{\mu }^{\lambda }R_{\lambda \nu }-R^{\sigma \lambda }R_{\lambda \mu \nu
\sigma }+\nabla _{\nu }\nabla _{\sigma }R_{\mu }^{\sigma }.  \label{1d}
\end{eqnarray}
When this is put in eq.$\left( \ref{1}\right) $ its left hand side becomes 
\begin{eqnarray*}
&&R_{\mu \nu }-\frac{1}{2}g_{\mu \nu }R+2a\left[ RR_{\mu \nu }-\frac{1}{4}%
g_{\mu \nu }R^{2}-\nabla _{\mu }\nabla _{\nu }R+g_{\mu \nu }\Box R\right] \\
&&+b\left[ -\frac{1}{2}g_{\mu \nu }R_{\lambda \sigma }R^{\lambda \sigma
}+\Box R_{\mu \nu }+\frac{1}{2}g_{\mu \nu }\Box R-\nabla _{\mu }\nabla _{\nu
}R+2R^{\sigma \lambda }R_{\lambda \mu \nu \sigma }\right] ,
\end{eqnarray*}

\noindent where I have taken into account the equalities 
\[
R^{\sigma \lambda }R_{\lambda \mu \nu \sigma }=R^{\sigma \lambda }R_{\nu
\sigma \lambda \mu }=R^{\lambda \sigma }R_{\lambda \nu \mu \sigma
}=R^{\sigma \lambda }R_{\lambda \nu \mu \sigma }, 
\]
and, taking into account that the divergence of Einstein\'{}s tensor eq.$%
\left( \ref{1c}\right) $ is zero, 
\[
\nabla _{\sigma }R^{\lambda \sigma }=\frac{1}{2}\nabla ^{\lambda }R,\;\nabla
_{\sigma }R_{\mu }^{\sigma }=\frac{1}{2}\nabla _{\mu }R. 
\]
The trace of the field equation is specially simple, namely 
\begin{equation}
(6a+2b)\Box R-R=kT_{\mu }^{\mu }\equiv kT.  \label{2}
\end{equation}

It is convenient to rewrite the field equation in terms of Einstein\'{}s
tensor, $G_{\mu \nu },$ and its trace, $G$, related to the Ricci tensor, $%
R_{\mu \nu },$ by 
\begin{equation}
R_{\mu \nu }=G_{\mu \nu }-\frac{1}{2}g_{\mu \nu }G,\;R=-G.  \label{1c}
\end{equation}
Also I shall write the field equation so that it looks like the standard GR
eq.$\left( \ref{E}\right) $, that is 
\begin{equation}
G_{\mu \nu }=kT_{\mu \nu }+kT_{\mu \nu }^{vac},  \label{4c}
\end{equation}
defining 
\begin{eqnarray}
T_{\mu \nu }^{vac} &\equiv &k^{-1}\{\left( 2a+b\right) \left[ g_{\mu \nu
}\Box G-\nabla _{\mu }\nabla _{\nu }G\right] +a\left[ GG_{\mu \nu }-\frac{1}{%
4}g_{\mu \nu }G^{2}\right]  \nonumber \\
&&+b\left[ 2G^{\sigma \lambda }R_{\lambda \mu \nu \sigma }-\frac{1}{2}g_{\mu
\nu }G_{\lambda \sigma }G^{\lambda \sigma }-\frac{1}{4}g_{\mu \nu
}G^{2}-\Box G_{\mu \nu }\right] \}.  \label{4d}
\end{eqnarray}
I stress that all results of this paper will be independent of whether we
assume that $T_{\mu \nu }^{vac}$ is a quantum vacuum stress-energy or we
consider eqs.$\left( \ref{4c}\right) $ and $\left( \ref{4d}\right) $
toghether as a modification of standard GR, maybe with no reference to the
vacuum. In any case I shall use throughout this paper a language appropriate
for the former assumption.

\section{Approximate vacuum field in a Newtonian star}

Our task is to solve the field equations for a spherically symmetric body,\
with mass $M$ and radius $R_{o}$, in a static spacetime with the condition
that the metric is asymptotically flat (Minkowskian). The body may be a
metallic sphere in a laboratory experiment, say like that of E\"{o}tv\"{o}s,
the earth or the sun, but to be specific I shall speak about a star from now
on. I will use standard (curvature) coordinates with metric 
\begin{equation}
ds^{2}=-\exp \left( \beta \left( r\right) \right) dt^{2}+\exp \left( \alpha
\left( r\right) \right) dr^{2}+r^{2}d\Omega ^{2}.  \label{9}
\end{equation}
In a static problem of spherical symmetry there are only 3 independent
components of Einstein\'{}s tensor eq.$\left( \ref{4c}\right) $, namely $%
G_{t}^{t},G_{r}^{r},G_{\theta }^{\theta }=G_{\phi }^{\phi },$ which are well
known functions of the metric parameters $\alpha $ and $\beta $\cite{Synge}$%
. $ Then we have 4 independent equations including a relation between
density and pressure; in particular for a star in equilibrium the latter
relation is the equation of state of matter, i, e, $p=p(\rho )$. We have
also 4 unknown functions, namely $\alpha \left( r\right) ,\beta \left(
r\right) ,\rho (r)$ and $p(r).$ The solution of these equations is involved
in general and here will be solved only for a few particular cases of
Newtonian, or slightly relativistic, stars.

Thus I shall consider stars where:

1${{}^{a}}$ The metric coefficient $\beta $, eq.$\left( \ref{9}\right) ,$ is
small compared with unity. In this case $\beta \simeq 2\Phi ,$ $\Phi $ being
the Newton potential which fulfils $\left| \Phi \right| \approx kM/R_{o}<<1$.

2${{}^a}$ The matter pressure, $p$, is small in comparison with the matter
density, $\rho .$ Actually this condition is related to the former because
we have $p/\rho \simeq kM/R_{o}.$ $\;$

In order to get the 3 components of $T_{\mu \nu }^{vac}(=k^{-1}G_{\mu \nu
}-T_{\mu \nu }^{vac}),$ we shall solve eq.$\left( \ref{4c}\right) $. For
Newtonian stars the equation may be approximated as follows. Firstly we may
neglect terms quadratic in $G_{\mu \nu }.$ In fact, the terms linear in $%
G_{\mu \nu }$ in the right side of eq.$\left( \ref{4d}\right) $ are of order 
$ak\rho /R_{\circ }^{2},$ $\rho $ being the typical matter density. In
contrast the terms quadratic in $G_{\mu \nu }$ are of order $ak^{2}\rho
^{2}, $ that is smaller than the former by $k\rho R_{\circ }^{2}\simeq $ $%
kM/R_{\circ }<<1.$ Thus eq.$\left( \ref{4c}\right) $ may be approximated by

\begin{equation}
G_{\mu \nu }-kT_{\mu \nu }\equiv kT_{\mu \nu }^{vac}\simeq \left(
2a+b\right) \left[ g_{\mu \nu }\Box G-\nabla _{\mu }\nabla _{\nu }G\right]
-b\Box G_{\mu \nu }.  \label{5}
\end{equation}
Secondly we may neglect the matter pressure in comparison with the density
in the interior of of the star, that is we may assume 
\begin{equation}
T_{t}^{t}=\rho \simeq T,\;T_{r}^{r}=T_{\theta }^{\theta }=T_{\phi }^{\phi
}=-p\simeq 0.  \label{4a}
\end{equation}
However I shall retain $p$ in some cases for the sake of clarity. Using the
metric eq.$\left( \ref{9}\right) $ we may write 
\[
\nabla _{\mu }\nabla _{\nu }G=\delta _{\mu }^{1}\delta _{\nu }^{1}\frac{%
d^{2}G}{dr^{2}}-\Gamma _{\mu \nu }^{1}\frac{dG}{dr}, 
\]
where I label $1$ the index of the radial coordinate, in order to avoid
confusion with the coordinate itself. The affine connections $\Gamma
_{tt}^{1}$ and $\Gamma _{11}^{1}$ are of order $kM/R_{\circ },$ whence the
terms involving them may be neglected. To the same order we may approximate 
\begin{equation}
g^{\theta \theta }\Gamma _{\theta \theta }^{1}=g^{\phi \phi }\Gamma _{\phi
\phi }^{1}\simeq -1/r,\;g^{11}\simeq 1,\;\Box G\simeq \nabla ^{2}G.
\label{9f}
\end{equation}
The term $\;\Box G_{\mu \nu }$ is more involved, although straightforward,
and I shall not write it in general. For our case, that is with the metric
eq.$\left( \ref{9}\right) $ and the approximations eqs.$\left( \ref{9f}%
\right) ,$ the 3 independent components of the tensor eq.$\left( \ref{5}%
\right) $ become 
\begin{eqnarray}
b\nabla ^{2}G_{t}^{t}+G_{t}^{t} &=&\left( 2a+b\right) \nabla ^{2}G+k\rho ,\;
\nonumber \\
b\nabla ^{2}G_{r}^{r}+\frac{4b}{r^{2}}(G_{\theta }^{\theta
}-G_{r}^{r})+G_{r}^{r} &=&\left( 2a+b\right) \frac{2}{r}\frac{dG}{dr}-kp, 
\nonumber \\
b\nabla ^{2}G_{\theta }^{\theta }+\frac{2b}{r^{2}}(G_{r}^{r}-G_{\theta
}^{\theta })+G_{\theta }^{\theta } &=&\left( 2a+b\right) (\frac{d^{2}G}{%
dr^{2}}+\frac{1}{r}\frac{dG}{dr})-kp.  \label{8}
\end{eqnarray}

At this moment I point out that, at a difference with general relativity,
here local isotropy of the matter stresses does not imply isotropy of the
spatial part of the Ricci tensor. That is, although the Einstein eq.$\left( 
\ref{E}\right) $ obviously leads to the implication 
\[
T_{r}^{r}=T_{\theta }^{\theta }\Rightarrow G_{r}^{r}=G_{\theta }^{\theta
}\Rightarrow R_{r}^{r}=R_{\theta }^{\theta }, 
\]
this is no longer true for the more involved field eq.$\left( \ref{4d}%
\right) $ (or the approximate eq.$\left( \ref{5}\right) .$) Indeed we may
have $G_{r}^{r}\neq G_{\theta }^{\theta }$ (and therefore anisotropy of the
vacuum stresses, i.e. $\left( T^{vac}\right) _{r}^{r}\neq \left(
T^{vac}\right) _{\theta }^{\theta }$ $)$ even if $T_{r}^{r}=T_{\theta
}^{\theta }.$ Furthermore the anisotropy of the gravitational field, i. e.
the inequality $G_{r}^{r}\neq G_{\theta }^{\theta }$, might induce
anisotropy of the matter stresses, i. e. $T_{r}^{r}\neq T_{\theta }^{\theta
} $ , but I shall not consider that possibility in this paper. Nevertheless
the matter local anisotropy, $T_{r}^{r}\neq T_{\theta }^{\theta },$ induced
by gravitational field anisotropy, $G_{r}^{r}\neq G_{\theta }^{\theta },$
might be relevant in strong gravitational fields.

\section{The field outside the star}

The solution of the field equations outside a spherical star, with a general 
$f(R),$ has been studied by several authors \cite{Chiba}. The interest of
the problem is that it puts constraints on the functions $f(R)$ in order to
be compatible with known facts in the solar system. Here I shall make a
similar calculation for our action eq.$\left( \ref{0a}\right) $ with the
purpose of finding the range of values of the parameters $a$ and $b$
compatible with terrestrial and solar system observations.

Thus our aim is to get the Einstein tensor, $G_{\mu }^{\nu },$ outside the
star taking into account that $T_{\mu }^{\nu }=0$ there. Hence the vacuum
stress-energy might be easily obtained, that is 
\[
\rho ^{vac}=k^{-1}G_{t}^{t},\;p_{r}^{vac}=-k^{-1}G_{1}^{1},\;p_{\theta
}^{vac}=-k^{-1}G_{\theta }^{\theta }. 
\]
In order to solve eqs.$\left( \ref{8}\right) $ I begin obtaining appropriate
linear combinations of them. If we add the first equation plus the second
one plus two times the third, we get the trace equation (compare with eq.$%
\left( \ref{2}\right) $) 
\begin{equation}
G-\left( 6a+2b\right) \nabla ^{2}\rho =kT.  \label{4b}
\end{equation}
where I have approximated $\Box T$ by the flat-space Laplacian of the matter
density, $\nabla ^{2}\rho $ (see eqs.$\left( \ref{4a}\right) $ and $\left( 
\ref{9f}\right) ).$ I shall assume $6a+2b>0,$ that is I exclude the case $%
6a+2b<0$ which, leading to an oscillating function $G(r)$, is unphysical.
The limiting case $6a+2b=0$ will be considered below.

Eq.$\left( \ref{4b}\right) $ in flat space may be solved by Green\'{}s
function method as follows. In order to simplify the writing I shall
sometimes use a dimensionless position vector, $\mathbf{x,}$ and a
dimensionless star radius $X$ defined by 
\begin{equation}
\mathbf{x}=\frac{\mathbf{r}}{\sqrt{6a+2b}},\;x=\left| \mathbf{x}\right| ,\;X=%
\frac{R_{o}}{\sqrt{6a+2b}},\;\gamma =\sqrt{\frac{6a+2b}{\left| b\right| }},
\label{x}
\end{equation}
where I have introduced also the parameter $\gamma $ for latter convenience.
Thus the fundamental solution for the trace eq.$\left( \ref{4b}\right) $ may
be written 
\begin{equation}
\nabla ^{2}f\left( \mathbf{x}\right) -f\left( \mathbf{x}\right) =-4\pi
\delta ^{3}\left( \mathbf{x}\right) \Rightarrow f=\frac{1}{x}\exp \left(
-x\right) ,  \label{a2}
\end{equation}
Hence the Einstein tensor outside the star may be obtained by integration,
giving 
\begin{eqnarray}
G_{t}^{t}+G_{r}^{r}+2G_{\theta }^{\theta } &\equiv &G\left( x\right) =\frac{k%
}{4\pi }\int_{\left| \mathbf{z}\right| <X}\frac{\rho \left( z\right) d^{3}%
\mathbf{z}}{\left| \mathbf{x-z}\right| }\exp \left( -\left| \mathbf{x-z}%
\right| \right)  \nonumber \\
&=&\frac{kM^{*}}{4\pi \left( 6a+2b\right) r}\exp \left( -\frac{r}{\sqrt{6a+2b%
}}\right) ,  \label{9c}
\end{eqnarray}
where 
\begin{equation}
M^{*}\equiv \left( 6a+2b\right) ^{3/2}\int_{0}^{X}\rho (z)4\pi z\sinh z\text{
}dz,\;  \label{a3}
\end{equation}
We see that for (small) objects fulfilling $R_{o}<<\sqrt{6a+2b},$ that is $%
X<<1,$ $\sinh z\simeq z,$ leading to $M^{*}\simeq M.$ This is not the case
for bodies such that $R_{o}\gtrsim \sqrt{6a+2b}$.

The physics behind these result may be better understood if we define a new
mass parameter 
\begin{equation}
M_{a}\equiv M^{*}\exp \left( -X\right) \equiv \left( 6a+2b\right) ^{3/2}\exp
\left( -X\right) \int_{0}^{X}\rho (z)4\pi z\sinh z\text{ }dz,  \label{a4}
\end{equation}
so that eq.$\left( \ref{9c}\right) $ may be rewritten 
\begin{equation}
G\left( r\right) =\frac{kM_{a}}{4\pi \left( 6a+2b\right) r}\exp \left( -%
\frac{r-R_{o}}{\sqrt{6a+2b}}\right) .  \label{trace}
\end{equation}
We see that the vacuum correction eq.$\left( \ref{trace}\right) $ may be
interpreted as if it depends on the distance, $r-R_{o},$ from the point $%
\mathbf{r}$\textbf{\ }where we measure $G$ to the closest point in the
surface of the star and the correction is due to the mass, $M_{a},$
contained in some volume of the star most close to the point \textbf{r}.
Indeed, the factor $\exp \left( -X\right) \sinh z$ in the integral eq.$%
\left( \ref{a4}\right) $ effectively restricts the range of integration to a
region near the surface. In particular for a body of constant density, like
the earth, or more generally any rocky planet or satellite, eq.$\left( \ref
{a4}\right) $ gives 
\begin{equation}
\frac{M_{a}}{M}=\frac{3}{2X^{2}}\left[ 1+\exp \left( -2X\right) \right] -%
\frac{3}{2X^{3}}\left[ 1-\exp \left( -2X\right) \right] \simeq \frac{3\left(
6a+2b\right) }{2R_{0}^{2}}<<1,  \label{mas}
\end{equation}
the latter equality being valid for $6a+2b>>R_{0}^{2}.$ In a star, like the
sun, the ratio $M_{a}/M$ would be much smaller because the surface density
is smaller than on earth and the radius much bigger.

In order to proceed with the calculation of the Einstein tensor outside the
star I subtract from twice the first eq.$\left( \ref{8}\right) $ the second
one and twice the third. This gives 
\begin{equation}
b\nabla ^{2}\left( 2G_{t}^{t}-G_{r}^{r}-2G_{\theta }^{\theta }\right)
+2G_{t}^{t}-G_{r}^{r}-2G_{\theta }^{\theta }=2\rho ,  \label{9a}
\end{equation}
again neglecting $p<<\rho .$ The solution outside the star is (compare with
eq.$\left( \ref{trace}\right) )$%
\begin{equation}
2G_{t}^{t}-G_{r}^{r}-2G_{\theta }^{\theta }=\frac{kM_{b}}{2\pi \left(
-b\right) r}\exp \left( -\frac{r-R_{o}}{\sqrt{-b}}\right) ,  \label{9b}
\end{equation}
where 
\begin{equation}
M_{b}\equiv \left( -b\right) ^{3/2}\exp \left( -\gamma X\right)
\int_{0}^{\gamma X}\rho (z)\sinh z4\pi zdz.  \label{mb}
\end{equation}
I have assumed $b<0$ because a positive $b$ would lead to an unphysical
oscillating function. For a body with constant density this leads to 
\begin{equation}
\frac{M_{b}}{M}=\frac{3}{2\gamma ^{2}X^{2}}\left[ 1+\exp \left( -2\gamma
X\right) \right] -\frac{3}{2\gamma ^{3}X^{3}}\left[ 1-\exp \left( -2\gamma
X\right) \right] \simeq \frac{3\left| b\right| }{2R_{0}^{2}}<<1,  \label{mbs}
\end{equation}
the latter equality valid for $\left| b\right| <<R_{0}^{2}.$

From eqs.$\left( \ref{trace}\right) $ and $\left( \ref{9b}\right) $ we get
outside the star 
\begin{equation}
\rho ^{vac}=k^{-1}G_{t}^{t}=\frac{1}{12\pi r^{3}}\left[ M_{a}x^{2}\exp
\left( X-x\right) +2M_{b}\gamma ^{2}x^{2}\exp \left( \gamma X-\gamma
x\right) \right] ,  \label{10}
\end{equation}
\begin{eqnarray}
p_{r}^{vac}+2p_{\theta }^{vac} &=&-k^{-1}\left( G_{r}^{r}+2G_{\theta
}^{\theta }\right)  \nonumber \\
&=&\frac{1}{6\pi r^{3}}\left[ M_{b}\gamma ^{2}x^{2}\exp \left( \gamma
X-\gamma x\right) -M_{a}x^{2}\exp \left( -x\right) \right] .  \label{10a}
\end{eqnarray}

In order to obtain separately the two different pressures, $p_{r}^{vac}$ and 
$p_{\theta }^{vac},$ I proceed as follows. The vacuum stress-energy tensor
eq.$\left( \ref{4d}\right) $ is divergence-free as may be easily checked.
Actually this property is a consequence of deriving the field equations from
an action functional. As the Einstein tensor has also zero divergence, eq.$%
\left( \ref{4c}\right) $ shows that both the matter and the vacuum
stress-energy tensors are separately divergence-free. The divergence of the
vacuum tensor gives the relation 
\begin{equation}
\frac{dp_{r}^{vac}}{dr}+2\frac{p_{r}^{vac}-p_{\theta }^{vac}}{r}=-\beta
^{\prime }\left( \rho ^{vac}+p_{r}^{vac}\right) \simeq 0,  \label{10b}
\end{equation}
where $\beta ^{\prime }$ is the radial derivative of the metric coefficient $%
\beta .$ The second equality follows from the fact that, in Newtonian stars, 
$r\beta ^{\prime }$ is of order $kM/R_{o}<<1,$ whilst $\rho ^{vac}$, $%
p_{r}^{vac}$ and $p_{\theta }^{vac}$ have the same order. Hence we obtain 
\[
\frac{d}{dr}\left( r^{3}p_{r}^{vac}\right) \simeq r^{2}\left(
p_{r}^{vac}+2p_{\theta }^{vac}\right) , 
\]
which, taking eq.$\left( \ref{10a}\right) $ into account, gives 
\begin{equation}
p_{r}^{vac}=\frac{1}{6\pi r^{3}}\left[ M_{a}\left( x+1\right) \exp \left(
X-x\right) -M_{b}\left( \gamma x+1\right) \exp \left( \gamma X-\gamma
x\right) \right] ,  \label{10c}
\end{equation}
\begin{equation}
p_{\theta }^{vac}=\frac{1}{12\pi r^{3}}\left[ M_{b}\left( \gamma
^{2}x^{2}+\gamma x+1\right) \exp \left( \gamma X-\gamma x\right)
-M_{a}\left( x^{2}+x+1\right) \exp \left( X-x\right) \right] .  \label{10d}
\end{equation}
where the labels eqs.$\left( \ref{x}\right) $ have been used. An integration
constant has been fixed so that $r^{3}p_{r}^{vac}\rightarrow 0$ for $%
r\rightarrow \infty $ in order that the absolute value of $p_{r}$ never
surpases $\rho ,$ eq.$\left( \ref{10}\right) .$

The limiting case 
\[
3a+b=0,-b>0 
\]
cannot be studied by the procedure leading to eqs.$\left( \ref{10}\right) $
to $\left( \ref{14}\right) $ but its trace eq.$\left( \ref{4b}\right) $ is
rather simple, namely 
\[
G=kT\Rightarrow T^{vac}=0\text{, in particular }G=0\text{ outside the star.} 
\]
Also eqs.$\left( \ref{8}\right) $ are simple and we obtain outside the star
(here $M_{a}=M_{b})$%
\begin{eqnarray*}
\rho ^{vac} &=&k^{-1}G_{t}^{t}=\frac{M_{a}}{18\pi ar}\exp \left( -\frac{%
r-R_{o}}{\sqrt{3a}}\right) ,\; \\
p_{r}^{vac} &=&-k^{-1}G_{r}^{r}=-\frac{M_{a}}{6\pi r^{3}}\left( \frac{r}{%
\sqrt{3a}}+1\right) \exp \left( -\frac{r-R_{o}}{\sqrt{3a}}\right) , \\
p_{\theta }^{vac} &=&-k^{-1}G_{\theta }^{\theta }=\frac{M_{a}}{12\pi r^{3}}%
\left( \frac{r^{2}}{3a}+\frac{r}{\sqrt{3a}}+1\right) \exp \left( -\frac{%
r-R_{o}}{\sqrt{3a}}\right) .
\end{eqnarray*}
We see that these equations are the limit of eqs.$\left( \ref{10}\right) $
to $\left( \ref{10d}\right) $ when $b\rightarrow -3a<0$. It is interesting
that, in this case, the vacuum field looks like a radiation field because
its stress-energy tensor is traceless. However that radiation field is not
isotropic, that is $p_{r}^{vac}\neq p_{\theta }^{vac}$, $p_{r}^{vac}$ being
negative and $p_{\theta }^{vac}$ positive$.$

Also the case $b=0$ cannot be studied by the procedure leading to eqs.$%
\left( \ref{10}\right) $ to $\left( \ref{14}\right) $ but eqs.$\left( \ref{8}%
\right) $ may be easily solved taking the trace eq.$\left( \ref{trace}%
\right) $ into account. We get

\begin{equation}
k\rho ^{vac}\equiv G_{t}^{t}=\frac{1}{3}G=\frac{kM_{a}}{72\pi ar}\exp \left(
-\frac{r-R_{o}}{\sqrt{6a}}\right) \text{ },  \label{21}
\end{equation}
\begin{equation}
-kp_{r}^{vac}\equiv G_{r}^{r}=\frac{1}{3}G-2a\frac{d^{2}G}{dr^{2}}=-\frac{%
kM_{a}}{6\pi r^{3}}\left( 1+\frac{r}{\sqrt{6a}}\right) \exp \left( -\frac{%
r-R_{o}}{\sqrt{6a}}\right) ,  \label{21a}
\end{equation}
\begin{equation}
-kp_{\theta }^{vac}\equiv G_{\theta }^{\theta }=\frac{1}{3}G-\frac{2a}{r}%
\frac{dG}{dr}=\frac{kM_{a}}{12\pi r^{3}}\left( \frac{r^{2}}{6a}+\frac{r}{%
\sqrt{6a}}+1\right) \exp \left( -\frac{r-R_{o}}{\sqrt{6a}}\right) .
\label{21b}
\end{equation}
It is interesting that, in this case, the mean vacuum pressure is negative
fulfilling $p_{mean}^{vac}=-\frac{2}{3}\rho ^{vac}.$

Finally in the particular case 
\[
2a+b=0, 
\]
eqs.$\left( \ref{10}\right) $ to $\left( \ref{14}\right) $ hold true, but it
is interesting to write them explicitly because they are specially simple,
that is 
\[
G=G_{t}^{t}=k\rho ^{vac}=\frac{kM_{a}}{8\pi ar}\exp \left( -\frac{r-R_{o}}{%
\sqrt{2a}}\right) ,G_{r}^{r}=p_{r}^{vac}=G_{\theta }^{\theta }=p_{\theta
}^{vac}=0. 
\]

In summary the vacuum density and stresses outside the star, derived from
the action eqs.$\left( \ref{0}\right) ,$ are given by eq.$\left( \ref{10}%
\right) $ and $\left( \ref{10a}\right) $, or the appropriate limits, for any
values of the parameters $a$ and $b$ fulfilling 
\begin{equation}
3a\geq -b\geq 0.  \label{14}
\end{equation}
For values violating these inequalities the solution of the field equation
outside a Newtonian spherical body would give an unphysical oscillatory
behaviour. Of course the values $a=b=0$ correspond to general relativity
(without vacuum field.)

Now I shall calculate the coefficients $\alpha \left( r\right) $ and $\beta
\left( r\right) $ of the metric eq.$\left( \ref{9}\right) $ taking into
account the well known relations\cite{Synge} 
\begin{equation}
\exp \left( -\alpha \right) =1-\frac{1}{r}\int_{0}^{r}u^{2}G_{t}^{t}\left(
u\right) du,\;\frac{d\beta }{dr}\equiv \beta ^{\prime }=\frac{e^{\alpha }-1}{%
r}-re^{\alpha }G_{r}^{r}.  \label{alfa}
\end{equation}
From the former it is straightforward to obtain the function $\alpha \left(
r\right) $ outside the star. We get, taking eq.$\left( \ref{10}\right) $
into account, 
\begin{equation}
\exp \left( -\alpha \right) =1-\frac{kM}{4\pi r}\left( 1-\delta \alpha
\right) ,  \label{13a}
\end{equation}
where 
\begin{eqnarray}
\delta \alpha \left( x\right) &=&\frac{1}{3}\frac{M_{a}}{M}\left[ \left(
x+1\right) \exp \left( X-x\right) -\left( X+1\right) \right]  \nonumber \\
&&+\frac{2}{3}\frac{M_{b}}{M}\left[ \left( \gamma x+1\right) \exp \left(
\gamma X-\gamma x\right) -\left( \gamma X+1\right) \right] .  \label{13aa}
\end{eqnarray}
We see that the dimensionless function $\delta \alpha $ represents the
correction to the $GR$ (Schwarzschild) exterior solution of a spherical star.

In order to get the function $\beta ^{\prime }\left( r\right) $ I start
expanding the metric coefficient $\exp \alpha $ in powers of the
gravitational constant $k$, a parameter which may be considered small
because $kM/R_{o}<<1$. That is 
\[
\exp \alpha =1+\frac{kM}{4\pi r}\left( 1-\delta \alpha \right) +\frac{%
k^{2}M^{2}}{16\pi ^{2}r^{2}}\left( 1-\delta \alpha \right) ^{2}+O\left(
k^{3}\right) . 
\]
Inserting this in the second eq.$\left( \ref{alfa}\right) $ we obtain $\beta
^{\prime }$ as an expansion in powers of $k$, that is 
\begin{eqnarray}
\beta ^{\prime } &=&\frac{kM}{4\pi r^{2}}\left( 1-\delta \alpha +\frac{4\pi
r^{3}p_{r}^{vac}}{M}\right)  \nonumber \\
&&+\frac{k^{2}M^{2}}{16\pi ^{2}r^{3}}\left( 1-\delta \alpha \right) \left(
1-\delta \alpha +\frac{4\pi r^{3}p_{r}^{vac}}{M}\right) +O\left(
k^{3}\right) ,  \label{13b}
\end{eqnarray}
where $\delta \alpha $ was given in eq.$\left( \ref{13aa}\right) $ and 
\begin{eqnarray*}
1-\delta \alpha +\frac{4\pi r^{3}p_{r}^{vac}}{M} &=&\frac{M_{a}}{3M}\left[
\left( x+1\right) \exp \left( X-x\right) +\left( X+1\right) \right] \\
&&+\frac{M_{b}}{3M}\left[ -4\left( \gamma x+1\right) \exp \left( \gamma
X-\gamma x\right) +2\left( \gamma X+1\right) \right] .
\end{eqnarray*}
This leads to 
\begin{eqnarray}
\beta ^{\prime } &=&\frac{kM}{4\pi r^{2}}+\frac{kM_{a}}{12\pi r^{2}}\left[
\left( x+1\right) \exp \left( X-x\right) +\left( X+1\right) \right] 
\nonumber \\
&&+\frac{kM_{b}}{6\pi r^{2}}\left[ \left( \gamma X+1\right) -2\left( \gamma
x+1\right) \exp \left( \gamma X-\gamma x\right) \right] +O\left(
k^{2}\right) .  \label{13bb}
\end{eqnarray}
Getting the term proportional to $k^{2}$ is straighforward, but the
resulting expression is involved and I will not write it explicitly.

Hence the parameter $\beta $ may be easily obtained by means of an
integration of eq.$\left( \ref{13b}\right) $ with the condition $\beta
\left( r\right) \rightarrow 0$ when $r\rightarrow \infty $. The result may
be written 
\begin{equation}
\beta =-\frac{kM}{4\pi r}(1+\delta \beta _{1})-\frac{k^{2}M^{2}}{32\pi
^{2}r^{2}}(1+\delta \beta _{2})+O\left( k^{3}\right) ,  \label{13c}
\end{equation}
where the function $\delta \beta _{1}\left( r\right) $ is the correction to
the $GR$ (Schwarzschild) solution to lowest order in $k$. It is 
\begin{equation}
\delta \beta _{1}\equiv \frac{M_{a}}{3M}\left[ \left( X+1\right) +\exp
(X-x)\right] +\frac{M_{b}}{3M}\left[ 2\left( \gamma X+1\right) -4\exp
(\gamma X-\gamma x)\right] .  \label{13cc}
\end{equation}
The term $\delta \beta _{2}$ is involved and I will not write it explicitly..

Eq.$\left( \ref{13c}\right) $ to order $O(k)$ is, therefore, 
\begin{eqnarray}
\beta &\simeq &-\frac{kM}{4\pi r}\left[ 1+\frac{M_{a}}{3M}\left( X+1\right) +%
\frac{2M_{b}}{3M}\left( \gamma X+1\right) \right]  \nonumber \\
&&-\frac{kM_{a}}{12\pi r}\exp (X-x)+\frac{M_{b}}{3\pi r}\exp (\gamma
X-\gamma x).  \label{betafi}
\end{eqnarray}
As is well known the term linear in the gravitational constant, $k$, in the
expansion of the metric parameter $\beta $ equals twice the Newtonian
potential. Therefore eq.$\left( \ref{betafi}\right) $ shows that the theory
resting upon the action eq.$\left( \ref{0}\right) $ predicts: 1) a
correction to the mass appearing in the Newtonian potential (the term within
square bracket), and 2) two non-Newtonian potentials of Yukawa type, one of
them attractive and the other one repulsive.

\section{Constraints on the parameters \textit{a} and \textit{b}}

\smallskip It is known that in f(R)-gravity the coefficient of the term $%
R^{2}$ (noted $a$ in this paper) cannot be greater than a few square
milimeters in order not to contradict laboratory experiments\cite{Faraoni}.
However this is not the case in the generalized theory here studied. In
fact, let us compare the Newtonian field intensity, $g\equiv kM/\left( 8\pi
r^{2}\right) $ with the field intensity eq.$\left( \ref{13bb}\right) $
predicted by our theory, which corresponds to half the term linear in $k$ of
eq.$\left( \ref{13bb}\right) $ and I shall label $g+\delta g.$ We get 
\begin{eqnarray}
\frac{\delta g}{g} &=&\frac{M_{a}}{3M}\left[ \left( x+1\right) \exp \left(
X-x\right) +\left( X+1\right) \right]  \nonumber \\
&&+\frac{2M_{b}}{3M}\left[ \left( \gamma X+1\right) -2\left( \gamma
x+1\right) \exp \left( \gamma X-\gamma x\right) \right] .  \label{g}
\end{eqnarray}
In the most interesting case of a body with constant density we may put eqs.$%
\left( \ref{mas}\right) $ and $\left( \ref{mbs}\right) $ in eq.$\left( \ref
{g}\right) $ and we get 
\begin{eqnarray}
\frac{\delta g}{g} &=&\frac{X\cosh X-\sinh X}{X^{3}}\left[ \left( x+1\right)
\exp \left( -x\right) +\left( X+1\right) \exp \left( -X\right) \right] 
\nonumber \\
&&+2\frac{\gamma X\cosh \left( \gamma X\right) -\sinh \left( \gamma X\right) 
}{\gamma ^{3}X^{3}}\left[ \left( \gamma X+1\right) \exp \left( -\gamma
X\right) -2\left( \gamma x+1\right) \exp \left( -\gamma x\right) \right] .
\label{gc}
\end{eqnarray}
This expression is small for both small and large $X$. Indeed for $X$ $%
\approx $ $x<<1$ we have, expanding in powers, 
\begin{equation}
\frac{\delta g}{g}\simeq 4\gamma ^{2}x^{2}-x^{2}-2\gamma
^{2}X^{2}-X^{2}=\left( \frac{4}{\left| b\right| }-\frac{1}{6a+2b}\right)
r^{2}-\left( \frac{2}{\left| b\right| }+\frac{1}{6a+2b}\right) R_{o}^{2}.
\label{g1}
\end{equation}
For $X>>1$ and $x-X<<X$ (i.e. $r-R_{o}<<R_{o}$) we obtain 
\begin{eqnarray}
\frac{\delta g}{g} &\simeq &\frac{1+2\gamma ^{-1}}{2X}+\frac{x}{2X^{2}}\exp
\left( X-x\right) -\frac{2x}{\gamma X^{2}}\exp \left( \gamma X-\gamma
x\right)  \nonumber \\
&\simeq &\frac{\sqrt{6a+2b}+2\sqrt{\left| b\right| }}{R_{o}}+\frac{\sqrt{%
6a+2b}}{R_{o}}\exp \left( -\frac{r-R_{o}}{\sqrt{6a+2b}}\right) -\frac{4\sqrt{%
\left| b\right| }}{R_{o}}\exp \left( -\frac{r-R_{o}}{\sqrt{\left| b\right| }}%
\right) .  \label{g2}
\end{eqnarray}

It is a fact that precise measurements have been made only for the
gravitational field of objects with sizes of meters or smaller (e.g. the
E\"{o}tv\"{o}s experiment) or for celestial bodies, in particular the earth%
\cite{Will}. Thus if \textit{both }$\sqrt{a}$\textit{\ and }$\sqrt{\left|
b\right| }$ have values between hundred meters and a few kilometers both eqs.%
$\left( \ref{g1}\right) $ and $\left( \ref{g2}\right) $ predict violations
of Newtonian gravity of order $10^{-4}$ or less. The two parameters should
be large enough because if $\left| b\right| $ is small then the ratio $%
\delta g/g$ becomes large in eq.$\left( \ref{gc}\right) $ and the theory
here developed is refuted by laboratory experiments\cite{Will}. It is worth
to remember here that during the late eigthies of the XX Century there were
some claims about the existence of non-Newtonian gravity (the ``fifth
force''), although a reanalysis of the experiments has lead to a consensus
that several uncertainties were not taken into account and the experiments
are actually compatible with Newtonian gravity within errors\cite{Fischbach}%
. In any case I should mention that some of the experiments apparently
showed the existence of two non-Newtonian potentials of Yukawa type, one
attractive and the other one repulsive, as in our eq.$\left( \ref{g2}\right) 
$. For instance in a tower experiment Eckhardt et al.\cite{Eckhardt}
reported a non-Newtonian gravity in the form 
\[
\frac{\delta g_{a}}{g}=1+\varepsilon _{at}\exp \left( -\frac{r}{\lambda _{at}%
}\right) -\varepsilon _{rep}\exp \left( -\frac{r}{\lambda _{rep}}\right) , 
\]
with values 
\begin{equation}
\varepsilon _{rep}-\varepsilon _{at}\simeq 0.007,\;\varepsilon _{at}\gtrsim
0.03,\;\lambda _{rep}\sim \lambda _{at}\sim 100\text{ m}.  \label{f}
\end{equation}
These values violate our eq.$\left( \ref{g2}\right) $ which predicts 
\[
\varepsilon \approx \frac{\lambda }{2R_{0}}\approx \frac{\lambda }{10^{7}%
\text{m}},\text{ }\lambda \approx \sqrt{a}. 
\]
In constrast this equality is compatible with known bounds provided that $%
\sqrt{a}<50$ km and $\varepsilon <0.005$ (see Fig.1 of Ref.\cite{Fischbach}%
.) I conclude that tests of our theory in the earth surface would require
experiments with errors several orders smaller than those typical of ``fifth
force'' experiments.

In summary there are two ranges where the parameters are compatible with all
performed experiments on Newtonian gravity, namely either both $\sqrt{a}$
and $\sqrt{\left| b\right| }$ are less than a few millimeters or both have
values between hundred meters and a few kilometers. But I shall point out
that, if $\sqrt{a}$ and $\sqrt{\left| b\right| }$ are in the latter range,
the correction eq.$\left( \ref{gc}\right) $ may be quite important for mass
concentrations with sizes of a few kilometers, like mountains. This may lead
to experimental tests of the theory, but I shall not discuss them in this
paper.

Even if the discrepancies between Newtonian gravity and the predictions of
eq.$\left( \ref{13b}\right) $ are too small to be detected, it is
interesting to see whether the theory here developed predicts corrections to
Newtonian gravity in measurable violation of general relativity. The
standard comparison is made using an isotropic metric rather than eq.$\left( 
\ref{9}\right) .$ However it is possible to make the comparison also with
the latter metric\cite{Weinberg}. For a spherical body the coefficients of
the metric eq.$\left( \ref{9}\right) $ may be expanded in powers of the
gravitational constant, $k$, in the form 
\begin{equation}
\exp \alpha =1+\eta \frac{kM}{4\pi r}+...,\;\exp \beta =1-\lambda \frac{kM}{%
4\pi r}+(\zeta -\lambda \eta )\frac{k^{2}M^{2}}{32\pi ^{2}r^{2}}+...
\label{13r}
\end{equation}
General relativity predicts $\lambda =\zeta =\eta =1.$ Furthermore, recent
measurements\cite{gamma} give the bound $\eta /\lambda =1+(2.1\pm 2.3)\times
10^{-5}.$

In order to make the comparison with the GR prediction eq.$\left( \ref{13r}%
\right) $ we need the expansion 
\begin{eqnarray}
\exp \beta &=&1+\beta +\frac{1}{2}\beta ^{2}+O\left( k^{3}\right)  \nonumber
\\
&\simeq &1-\frac{kM}{4\pi r}(1+\delta \beta _{1})+\frac{k^{2}M^{2}}{32\pi
^{2}r^{2}}\left( 1+2\delta \beta _{1}+(\delta \beta _{1})^{2}-\delta \beta
_{2}\right) .  \label{13s}
\end{eqnarray}
Now the comparison of eq.$\left( \ref{13r}\right) $ with eqs.$\left( \ref
{13a}\right) $ and $\left( \ref{13s}\right) $ gives (to lowest order in $k$) 
\[
\lambda =1-\delta \alpha ,\eta =1+\delta \beta _{1},\zeta =1+3\delta \beta
_{1}+(\delta \beta _{1})^{2}-\delta \beta _{2}-\delta \alpha -\delta \alpha
\delta \beta _{1}. 
\]
With arguments similar to those used in the analysis of eq.$\left( \ref{g}%
\right) $ we get $\lambda =\eta =\zeta =1$ with errors smaller than $10^{-3%
\text{ }}$ for the Earth gravity and smaller than 10$^{-6}$ for the gravity
of the sun, provided that the parameters $\sqrt{a}$ and $\sqrt{\left|
b\right| }$ are less than a few kilometers. The reason for the difference
between terrestrial and solar gravity lies in that, according to our
predictions, the quantities $\delta \alpha ,\delta \beta _{1}$ and $\delta
\beta _{2}$ are proportional to the density near the surface (much smaller
in the sun ) and inversely proportional to the radius (100 times greater in
the sun). In particular I get 
\begin{equation}
\left| \frac{\eta }{\lambda }-1\right| \simeq \left| \frac{2M_{b}}{3M}\left(
\gamma x-1\right) \exp \left( \gamma X-\gamma x\right) +\frac{M_{a}}{3M}%
(x+2)\exp \left( X-x\right) \right| <2.10^{-5},  \label{13v}
\end{equation}
in agreement with observations\cite{gamma}.

For the gravitational interactions of bodies with sizes of a few meters or
smaller, used in laboratory experiments, the ratio $\eta /\lambda $ is of
order $R_{o}^{2}/a\leq 10^{-6}$ if $\sqrt{a}$ and $\sqrt{\left| b\right| }$
are larger than about one kilometer.

\section{Hydrostatic equilibrium}

The structure of a spherical star in equilibrium may be obtained from eqs.$%
\left( \ref{4c}\right) $ and $\left( \ref{4d}\right) $ plus the equation of
state. The solution of these equations is involved and furthermore the fact
that they are fourth-order shows that we need four (initial or boundary)
conditions, which may give rise to some ambiguity. I will not attempt
solving exactly those equations in this paper, but I shall sketch the
solution of the approximate eqs.$\left( \ref{8}\right) $ inside Newtonian or
slightly relativistic stars.

I shall begin solving the trace eq.$\left( \ref{4b}\right) ,$ via eq.$\left( 
\ref{9c}\right) ,$ in the interior of the star, that is $x<X$. I get 
\begin{equation}
G\left( x\right) =\frac{k}{2x}\int_{0}^{X}\rho \left( z\right) \left[ \exp
\left( -\left| x-z\right| \right) -\exp \left( -z-x\right) \right] zdz.
\label{a9}
\end{equation}
For a star where $X>>1$ we may make the following 3 approximations: 1)
neglecting the second exponential in comparison with the first one (except
near the origin), 2) approximating $\rho \left( z\right) $ by 
\[
\rho \left( z\right) \simeq \rho \left( x\right) +\left[ \frac{d\rho }{dz}%
\right] _{z=x}\left( z-x\right) +\frac{1}{2}\left[ \frac{d^{2}\rho }{dz^{2}}%
\right] _{z=x}\left( z-x\right) ^{2}, 
\]
and 3) extending the z-integration to the interval $\left( -\infty ,\infty
\right) .$ Then the integral is trivial and we get 
\begin{equation}
G\left( x\right) =k\left( T+T^{vac}\right) \simeq k\left( \rho \left(
x\right) +\frac{2}{x}\frac{d\rho }{dx}+\frac{d^{2}\rho }{dx^{2}}\right) .
\label{a10}
\end{equation}
This same result is obtained near the center although the approximations
involved should be different. It may be realized that the last two terms of
eq.$\left( \ref{a10}\right) $ are of order $X^{-2}=\left( 6a+2b\right)
/R_{o}^{2}<<1$ and the terms neglected are of the order of $X^{-4}.$

The same result may be obtained if we combine eqs.$\left( \ref{4b}\right) $
and $\left( \ref{4c}\right) ,$ the latter with $\nabla ^{2}$ substituted for 
$\Box ,$ and work to the same order of approximation$.$ In fact we get 
\begin{equation}
T^{vac}=\left( 6a+2b\right) \nabla ^{2}\left( T+T^{vac}\right) \simeq \left(
6a+2b\right) \nabla ^{2}T\simeq \left( 6a+2b\right) \nabla ^{2}\rho .
\label{a11}
\end{equation}
The agreement between both calculations, leading to eqs.$\left( \ref{a10}%
\right) $ and $\left( \ref{a11}\right) ,$ respectively, reinforces the
validity of solving eqs.$\left( \ref{8}\right) $ by approximating $G_{\mu
}^{\nu }$ by $kT_{\mu }^{\nu }$ in all terms which are linear in the small
parameters $a$ or $b$. I point out that this approximation was not valid in
the study, made in Section 4, of the exterior of the star because $T_{\mu
}^{\nu }=0$ there. Thus eqs.$\left( \ref{8}\right) $ lead, in the star
interior, to 
\begin{eqnarray}
\rho ^{vac} &\equiv &(T^{vac})_{t}^{t}\simeq 2a\left( \frac{2}{r}\frac{d\rho 
}{dr}+\frac{d^{2}\rho }{dr^{2}}\right) \equiv 2a\left( \frac{2}{r}\rho
^{\prime }+\rho ^{\prime \prime }\right) ,  \nonumber \\
p_{r}^{vac} &\equiv &-(T^{vac})_{r}^{r}\simeq -\left( 2a+b\right) \frac{2}{r}%
\rho ^{\prime },\;  \nonumber \\
p_{\theta }^{vac} &\equiv &-(T^{vac})_{\theta }^{\theta }\simeq -\left(
2a+b\right) \left( \frac{1}{r}\rho ^{\prime }+\rho ^{\prime \prime }\right) .
\label{7}
\end{eqnarray}
These results are consistent with eq.$\left( \ref{a11}\right) $ for the
trace $T^{vac}.$ It is remarkable that the vacuum density in the interior of
the star depends only on the parameter $a$, and not on $b$, within our
approximations (in particular neglecting the matter pressure in comparison
with the matter density). If $-b<2a$ the mean vacuum pressure is negative in
the interior of the star.

I shall point out that eqs.$\left( \ref{7}\right) $ include \textit{all
vacuum effects }to our order of approximation. In particular we should not
add to eq.$\left( \ref{7}\right) $ the effect of the Yukawa-type
contribution to the Newtonian potential (see eqs.$\left( \ref{13b}\right) $%
). It is remarkable that the density of the vacuum contribution is negative
in the central part of the star because both $d\rho /dr$ and $d^{2}\rho
/dr^{2}$ are negative there. However the total density is positive
everywhere when eq.$\left( \ref{7}\right) $ is valid because we have assumed
that $\left| \rho ^{vac}\right| <<\rho $ in deriving it$.$ The vacuum
density is positive near the surface of the star and the vacuum contribution
to the total mass is zero, that is 
\begin{equation}
m^{vac}=\int_{0}^{r}2a\left( \frac{2}{r}\frac{d\rho }{dr}+\frac{d^{2}\rho }{%
dr^{2}}\right) 4\pi r^{2}dr=8\pi ar^{2}\frac{d\rho }{dr}\leq
0,\;M^{vac}=m^{vac}\left( R_{o}\right) =0,  \label{7m}
\end{equation}
provided that $d\rho /dr$ is zero at the star surface. In particular this is
the case in all polytropes with $\gamma <2$ (see next Section.) I stress
that all these properties are valid only for large spherical static bodies,
where eqs.$\left( \ref{7}\right) $ are a good enough approximation. I shall
point out that, strictly speaking, the total mass of the star should include
the mass of the vacuum contribution outside the star (with density given by
eq.$\left( \ref{10}\right) )$ so that $M^{vac}$ as defined in eq.$\left( \ref
{7m}\right) $ is only the mass in the interior of the star. The external
mass due to the vacuum may be obtained by integration of the density eq.$%
\left( \ref{10}\right) $ giving 
\begin{equation}
M_{ext}^{vac}\simeq \frac{R_{o}}{3}\left( \frac{M_{a}}{\sqrt{6a+2b}}+2\frac{%
M_{b}}{\sqrt{-b}}\right) \simeq \frac{M}{R_{o}}\left( \sqrt{6a+2b}+2\sqrt{-b}%
\right) ,  \label{mext}
\end{equation}
the latter equality being valid for a celestial body of constant density
like the earth.

The three independent components of the Einstein-type eq.$\left( \ref{4c}%
\right) $ become 
\begin{eqnarray}
G_{t}^{t} &=&k\rho +2ka\left( \frac{2}{r}\rho ^{\prime }+\rho ^{\prime
\prime }\right) ,\;G_{r}^{r}=-kp+\left( 2a+b\right) k\frac{2}{r}\rho
^{\prime },  \nonumber \\
G_{\theta }^{\theta } &=&-kp+\left( 2a+b\right) k\left( \frac{1}{r}\rho
^{\prime }+\rho ^{\prime \prime }\right) ,  \label{7s}
\end{eqnarray}
where $\rho ^{\prime }=d\rho /dr,\rho ^{\prime \prime }=d^{2}\rho /dr^{2}.$
Eqs.$\left( \ref{7s}\right) $ will be the starting point for all
calculations of structure of stars to be made in the following.

The condition that Einstein\'{}s tensor $G_{\mu }^{\nu }$ is divergence-free
leads to the following hydrostatic equilibrium equation\cite{Herrera} 
\begin{equation}
\frac{dp_{r}^{eff}}{dr}+\frac{2}{r}\left( p_{r}^{eff}-p_{\theta
}^{eff}\right) =-\frac{k}{2}\frac{m^{eff}+4\pi r^{3}p_{r}^{eff}}{4\pi
r^{2}-krm^{eff}}\left( \rho ^{eff}+p_{r}^{eff}\right) ,  \label{52}
\end{equation}
where I define 
\begin{eqnarray}
\rho ^{eff} &=&\rho +\rho ^{vac},\;p_{r}^{eff}=p+p_{r}^{vac},\;p_{\theta
}^{eff}=p+p_{\theta }^{vac},\;  \nonumber \\
m^{eff} &=&4\pi \int_{0}^{r}r^{2}\rho ^{eff}dr.  \label{53}
\end{eqnarray}
Putting eqs.$\left( \ref{7}\right) $ into eq. $\left( \ref{52}\right) $ we
get

\begin{equation}
\frac{dp}{dr}=-\frac{k}{2}\frac{m+4\pi r^{3}p-8\pi \left( a+b\right) r^{2}%
\frac{d\rho }{dr}}{4\pi r^{2}-krm-8\pi akr^{3}\frac{d\rho }{dr}}\left( \rho
+p-2b\frac{1}{r}\frac{d\rho }{dr}+2a\frac{d^{2}\rho }{dr^{2}}\right) .
\label{7c}
\end{equation}
I point out again that this equilibrium equation might be seen as either a
modification (to first order in $a$ and $b$) of the general relativistic
equilibrium equation, due to the change from the Einstein-Hilbert action to
the action eq.$\left( \ref{0a}\right) $, or as an effect of the quantum
vacuum. Both interpretations lead to the same physical consequences.

In going from eq.$\left( \ref{52}\right) $ to eq.$\left( \ref{7c}\right) $
it is interesting the cancellation 
\begin{equation}
\frac{dp_{r}^{vac}}{dr}+\frac{2}{r}\left( p_{r}^{vac}-p_{\theta
}^{vac}\right) =\frac{dp_{r}^{vac}}{dr}+\frac{2}{r}\left(
p_{r}^{eff}-p_{\theta }^{eff}\right) =0,  \label{7p}
\end{equation}
which actually derives from eq.$\left( \ref{10b}\right) .$ A fortunate
consequence of this cancelation is that eq.$\left( \ref{7c}\right) $ is an
ordinary differential equation of \textit{third order} (in the variable $%
m(r) $ ), rather than fourth order, when $p$ is written in terms of $\rho ,$
e. g. via the equation of state. Then we may use the initial conditions $%
m\left( 0\right) =0$, $\left[ d\rho /dr\right] _{r=0}=0$ and only one more
condition is needed, e. g. fixing the central density, $\rho \left( 0\right)
.$ Thus the solutions of eq.$\left( \ref{7c}\right) $ consist of a
one-parameter family as in the standard theory of stars in equilibrium.

Assuming that both the relativistic and the vacuum corrrections are small,
eq.$\left( \ref{7c}\right) $ may be approximated as 
\begin{eqnarray}
\frac{dp}{dr} &=&-\frac{km}{8\pi r^{2}}\rho \left( 1+GR+vacuum\right) ,
\label{55} \\
GR &\equiv &\frac{4\pi r^{3}p}{m}+\frac{km}{4\pi r}+\frac{p}{\rho }, 
\nonumber \\
vacuum &\equiv &-\left( a+b\right) \frac{8\pi r^{2}}{m}\frac{d\rho }{dr}-2b%
\frac{1}{r\rho }\frac{d\rho }{dr}+2a\frac{1}{\rho }\frac{d^{2}\rho }{dr^{2}}.
\nonumber
\end{eqnarray}
It may be realized that the first factor on the right side represents the
Newtonian approximation, the 3 terms labelled $GR$ are the well known first
order corrections of general relativity and those labelled $vacuum$ are
corrections due to the vacuum stress-energy. The former (latter) are of
order $kM/R$ (order $a/R^{2}$) with respect to the Newtonian approximation.
I have ignored terms which are both vacuum and GR corrections, that is of
order $akM/R^{3}$ with respect to the Newtonian approximation. As may be
seen, and it is well known, the three GR corrections are positive, that is
every one contributes to the increase of gravitational effects. In contrast
the second term of $vacuum$ is negative (remember that $b<0$ and $d\rho
/dr<0 $) and the third one is negative in the central region of the star
because $d^{2}\rho /dr^{2}$ is negative there. The first vacuum term would
be positive (negative) everywhere if $\left| b\right| <a$ $\left( \left|
b\right| >a\right) .$ Finally the third term is positive near the surface.
Consequently from eq.$\left( \ref{55}\right) $ no conclusion seems possible
about whether the vacuum corrections increase or decrease the effects of
gravity in comparison with the Newtonian approximation.

There is, however, a simple argument which gives the answer. The
gravitational interaction energy, in the Newtonian approximation but with
the vacuum correction included, is given by 
\begin{eqnarray}
V &=&-\frac{k}{8\pi }\int d^{3}\mathbf{r}_{1}\rho ^{eff}\left( \mathbf{r}%
_{1}\right) \int d^{3}\mathbf{r}_{2}\rho ^{eff}\left( \mathbf{r}_{2}\right)
\left| \mathbf{r}_{1}-\mathbf{r}_{2}\right| ^{-1}  \nonumber \\
&\simeq &V_{matter}-\frac{k}{4\pi }\int d^{3}\mathbf{r}_{1}\rho \left( 
\mathbf{r}_{1}\right) \int d^{3}\mathbf{r}_{2}\rho ^{vac}\left( \mathbf{r}%
_{2}\right) \left| \mathbf{r}_{1}-\mathbf{r}_{2}\right| ^{-1}  \nonumber \\
&=&V_{matter}-k\int_{0}^{R_{o}}\rho \left( r\right) m\left( r\right)
^{vac}rdr=V_{matter}+3ka\int_{0}^{R_{o}}\rho ^{2}4\pi r^{2}dr,  \label{evac}
\end{eqnarray}
where we have taken eq.$\left( \ref{7m}\right) $ into account and neglected
the vacuum density in the exterior of the star (but see below). Also I have
performed an integration by parts in the second equality assuming that $\rho
=0$ at the star\'{}s surface. The reason why the vacuum correction is
positive derives from the fact that $m^{vac}(r)\leq 0$ (see eq.$\left( \ref
{7m}\right) )$ so that the vacuum contribution makes the Newtonian
gravitational interacction energy \textit{less negative}. It is as if there
were a short range repulsion given by 
\[
3ka\delta ^{3}\left( \mathbf{r}_{1}-\mathbf{r}_{2}\right) . 
\]
Actually the vacuum density outside the star gives another contribution to
the gravitational interaction which is negative. In fact, taking eq.$\left( 
\ref{10}\right) $ into account we have 
\begin{eqnarray}
V_{ext}^{vac} &\simeq &-\frac{R_{o}}{3}\left( \frac{M_{a}}{\sqrt{6a+2b}}+2%
\frac{M_{b}}{\sqrt{-b}}\right) \frac{kM}{8\pi R_{o}}  \nonumber \\
&\simeq &-\frac{M}{R_{o}}\left( \sqrt{6a+2b}+2\sqrt{-b}\right) \frac{kM}{%
8\pi R_{o}},  \label{evacex}
\end{eqnarray}
the latter equality being valid for a celestial body of constant density
like the earth. Considering the two terms of eq.$\left( \ref{evac}\right) $
plus the terms eqs.$\left( \ref{mext}\right) $ and $\left( \ref{evacex}%
\right) $ we conclude the following. For the earth the Newtonian
gravitational energy, $V,$ and the vacuum term, $M_{ext}^{vac},$ may be of
similar size (their ratio is of order $kM/\sqrt{a}),$ whilst the other two
terms are much smaller. However in star, where the surface density is very
small, the dominant term may be $V$, followed by $V^{vac},$ whilst the other
two terms are much smaller than the latter (for the proof in the case of
white dwarfs see next Section).

The relative importance of the vacuum and GR corrections may be estimated
from the ratio of the first vacuum correction to the first GR correction in
eq.$\left( \ref{55}\right) .$ It is 
\begin{equation}
\frac{vacuum}{GR}\sim \frac{2a+b}{rp}\frac{d\rho }{dr}\sim \frac{a}{R^{2}}%
\times \frac{\rho }{p}.  \label{56}
\end{equation}
In Newtonian stars the first factor on the right side is small, as was shown
in the second Section, and the second factor is large. Thus no conclusion
may be reached (in any case both GR and vacuum corrections are small). In
compact objects like neutron stars the second factor is of order unity and
the first one might be large, which suggests that the vacuum corrections
could be very relevant. However in these stars many of the approximations
leading to eq.$\left( \ref{55}\right) $ are not valid and a calculation
starting from eqs.$\left( \ref{4c}\right) $ and $\left( \ref{4d}\right) $
would be necessary.

As an illustrative example in the following I study the effect of the vacuum
corrections on the internal properties of the sun. I shall consider, in
particular, the change in central pressure due to the correction terms given
in eq.$\left( \ref{55}\right) .$ The change in the central pressure will be $%
\delta p\left( 0\right) $ where 
\begin{eqnarray}
\delta p\left( r\right) &=&-\int_{r}^{R_{\odot }}\frac{d\delta p}{dr}dr=%
\frac{k}{8\pi }\int_{r}^{R_{\odot }}\frac{m}{r^{2}}\rho dr\left[
vacuum\right]  \label{56a} \\
&=&-\left( a+b\right) k\int_{r}^{R_{\odot }}\rho \frac{d\rho }{dr}%
dr-bk\int_{r}^{R_{\odot }}\frac{m}{4\pi r^{3}}\frac{d\rho }{dr}%
dr+ak\int_{r}^{R_{\odot }}\frac{m}{4\pi r^{2}}\frac{d^{2}\rho }{dr^{2}}dr. 
\nonumber
\end{eqnarray}
An integration by parts of the third term gives an integrated part, a term
which combines with the first one and can be integrated easily, and a new
term which combines with the second. Thus we have (assuming that $\rho
^{\prime }\left( R_{o}\right) =0)$%
\begin{equation}
\delta p\left( 0\right) =\left( a+\frac{b}{2}\right) k\left[ \rho \left(
0\right) \right] ^{2}+\left( 2a-b\right) k\int_{0}^{R_{\odot }}\frac{m}{4\pi
r^{3}}\frac{d\rho }{dr}dr.  \label{56aa}
\end{equation}
The integral in the second term may be performed numerically using data of
the calculations made by Bahcall et al.\cite{Bahcall}, but for our purposes
it is enough to find an upper bound. Taking into acount that $m/r^{3}\leq 
\frac{4}{3}\pi \rho \left( 0\right) $ and $d\rho /dr<0,$ the value of the
second integral lies in the interval $\left( -\frac{1}{3}\left[ \rho \left(
0\right) \right] ^{2},0\right) .$ Thus we get 
\[
\left( \frac{1}{3}a+\frac{5}{6}b\right) k\left[ \rho \left( 0\right) \right]
^{2}\leq \delta p\left( 0\right) \leq \left( a+\frac{b}{2}\right) k\left[
\rho \left( 0\right) \right] ^{2}, 
\]
which shows that 
\[
\left| \delta p\left( 0\right) \right| \approx (10\text{ km)}^{2}\times
7.4\times 10^{-29}\text{cm/g}\times \left( 156\text{g/cm}^{3}\right)
^{2}=1.8\times 10^{-12}\text{ g/cm}^{3} 
\]
where I have estimated the parameters $a$ and $\left| b\right| $ in about 10
km. We see that $\delta p\left( 0\right) $ may be negative (e.g. for $\left|
b\right| >2a)$ and positive (e.g. for $\left| b\right| <2a/5.)$ The central
pressure is (in units of mass/volume) 
\[
p=1.19\frac{\rho k_{B}\Theta }{m_{H}c^{2}}=2.6\times 10^{-4}\text{ g/cm}%
^{3}, 
\]
where $k_{B}$ is Boltzmann\'{}s constant, $m_{H}$ the mass of the hydrogen
atom and $\Theta $ the the temperature. The factor $1.19$ is the mean number
of particles per baryon, which derives from the chemical composition in
center of the sun, that is 35\% hydrogen and 65\% helium. Hence we get 
\begin{equation}
\frac{\delta p\left( 0\right) }{p\left( 0\right) }=\frac{\delta \Theta
\left( 0\right) }{\Theta \left( 0\right) }\approx 10^{-8}.  \label{70}
\end{equation}
We see that the effect of the vacuum stress-energy in the structure of the
sun is negligible. In particular it gives no change in the prediction of
neutrino emission from the center of the sun.

\section{Stability of $\gamma =4/3$ polytropes. Application to white dwarfs}

In the following I shall apply the theory resting upon the Einstein-type eqs.%
$\left( \ref{7s}\right) $ to the study of equilibrium and stability of $%
\gamma =4/3$ polytropes, that is Newtonian stars with equation of state of
the form $p=K\rho ^{\gamma },\gamma =4/3.$ In order to make more transparent
the comparison with the literature, in this section I shall use Newton\'{}s
constant, $G$, rather than $k=8\pi G.$ No confussion should arise with the
Einstein tensor, which will not be mentioned in this section.

It is well known that polytropes are stable if $\gamma >4/3$ and unstable if 
$\gamma <4/3.$ If $\gamma =4/3$ the stability depends on small corrections
which therefore become relevant. In particular, general relativistic
corrections produce unstability, although other corrections may compensate
for that. Stars which may be treated as $\gamma =4/3$ polytropes are white
dwarfs and supermassive stars. Both are stable when the internal energy is
large enough, but become unstable after radiating a fraction of that energy%
\cite{Shapiro}. In both cases it is assumed that a source of unstability are
the relativistic corrections, although in some white dwarfs also
neutronization of the core may induce unstabilility. Here I shall study
white dwarfs but not supermassive stars. Indeed the latter are hypothetical
and there are no observations on them. In addition the possible corrections
due to the quantum vacuum, as they are proposed in this paper, are probably
small because typical dimensions of supermassive stars are far larger than
the parameters $\sqrt{a}$ and $\sqrt{-b}.$

White dwarfs are formed from ordinary stars after a period of cooling. The
theory here considered corresponds to stars sufficiently cold to be treated
as if the temperature is zero Kelvin. Also I will consider only massive
white dwarfs because small ones are well approximated by $\gamma =5/3$
polytropes$,$ they are always stable and have no interest here. The theory
of white dwars in our approach might consists of solving the hydrostatic
equilibrium eq.$\left( \ref{7c}\right) $ with the equation of state 
\begin{equation}
\rho =m_{H}n+u(n)\Rightarrow p=n\frac{d\rho }{dn}-\rho =n\frac{du}{dn},
\label{eos}
\end{equation}
$n$ being the baryon density, $\rho $ the mass density and $p$ the pressure.
The mass $m_{H}$ is close to that of the hydrogen atom (its precise value
depends on the chemical composition of the star.) The function $u\left(
n\right) $ corresponds to a $\gamma =4/3$ polytriope with small corrections,
that is 
\begin{equation}
u(n)=3K(m_{H}n)^{4/3}+\delta u(n)\Rightarrow p\simeq K\rho ^{4/3},\;u\simeq
3K\rho ^{4/3}  \label{eos1}
\end{equation}
where $K$ is a constant. In the Newtonian approximation eqs.$\left( \ref{7c}%
\right) $ and $\left( \ref{eos}\right) $ become 
\begin{equation}
\frac{dp}{dr}=-\frac{Gm}{r^{2}}\rho ,\;p=K\rho ^{4/3}\Rightarrow \frac{d\rho 
}{dr}=-\frac{3}{4K}\frac{Gm(r)}{r^{2}}\rho (r)^{2/3}.  \label{N1}
\end{equation}

If we introduce in eq.$\left( \ref{N1}\right) $ the appropriate corrections
to lowest order we get 
\begin{equation}
\frac{d\rho }{dr}=-\frac{3}{4K}\frac{Gm(r)}{r^{2}}\rho (r)^{2/3}\left(
1+GR+vacuum+eos\right) ,  \label{N2}
\end{equation}
where $GR$ and $vacuum$ were given in eq.$\left( \ref{55}\right) .$ The
additional term, $eos$, corresponds to the modification, $\delta u(n),$ of
the polytropic equation of state mentioned in eq.$\left( \ref{eos1}\right) ,$
which I will not write explicitly here. Only the central density, $\rho
\left( 0\right) \equiv \rho _{c},$ is needed in order to fix one solution of
eq.$\left( \ref{N2}\right) $ (see comment after eq.$\left( \ref{7p}\right)
). $ Thus for every value of $\rho _{c}$ the solution of eq.$\left( \ref{N2}%
\right) $ provides the functions $\rho \left( r\right) ,$ $p\left( r\right) $
and $n(r).$ Hence we could calculate the baryon number, $N$, and the mass of
the star, $M^{eff}$, as functions of $\rho _{c}$, using the well known
expressions

\begin{equation}
N\equiv \int_{0}^{R_{o}}\frac{m_{H}n\left( r\right) 4\pi r^{2}dr}{\sqrt{%
1-2Gm^{eff}\left( r\right) /r}},\;m^{eff}\left( r\right) \equiv
\int_{0}^{r}4\pi x^{2}dx\rho ^{eff}\left( x\right) ,\;M^{eff}=m^{eff}\left(
\infty \right) ,  \label{57e}
\end{equation}
$R_{o}$ being the radius of the star and $\rho ^{eff}$ the sum of the matter
and vacuum densities (see eq.$\left( \ref{53}\right) $.) The upper limit of
the integral giving the total mass, $M^{eff}$, is $\infty ,$ rather than $%
R_{o}$, because we should include the mass associated to the vacuum density
both inside and in the neighbourhood of the star (see eq.$\left( \ref{mext}%
\right) .)$ Actually the quantity of interest is the binding energy defined
by 
\begin{equation}
E\equiv M^{eff}-m_{H}N.  \label{N22}
\end{equation}
which for Newtonian stars (i.e. without the corrections $GR,vacuum$ and $%
eos) $ becomes 
\begin{equation}
E=3K\int_{0}^{R_{o}}\rho ^{4/3}4\pi r^{2}dr-G\int_{0}^{R_{o}}m\rho 4\pi rdr.
\label{BE}
\end{equation}
In summary eqs.$\left( \ref{N2}\right) $ and $\left( \ref{N22}\right) $
provide a one-parameter family of \textit{solutions of equilibrium, }and the
question is which of such solutions correspond to stable equilibrium.

A standard method to study stability is to start from an equilibrium
configuration and perform the transformation $r\rightarrow \lambda r,\rho
\rightarrow \lambda ^{-3}\rho $ which leaves the mass unchanged, that is $%
M\rightarrow M.$ The binding energy, $E$, becomes a function of $\lambda $
and the configuration is stable if the function $E\left( \lambda \right) $
has a minimum for $\lambda =1.$ Thus the equilibrium and stability of white
dwarfs may be studied as follows. The solution of the Newtonian eq.$\left( 
\ref{N1}\right) $ for a central density $\rho _{c}$ provides the density $%
\rho \left( \rho _{c};r\right) $ as a function of the radial coordinate. We
may assume that this function is a good approximation for the solution of eq.%
$\left( \ref{N2}\right) $ with the said central density $\rho _{c}.$ Thus we
may use it in order to get the baryon number, $N$, and the binding energy, $%
E $, via eqs.$\left( \ref{57e}\right) $ and $\left( \ref{N22}\right) ,$ as
functions of the central density$.$ Now we consider star configurations out
of equilibrium, but close to the one given by the function $\rho \left( \rho
_{c};r\right) ,$ by performing the $\lambda $ transformation above stated.
Thus we get the binding energy, $E\left( \rho _{c},\lambda \right) $, and
the baryon number, $N\left( \rho _{c},\lambda \right) $, as functions of $%
\rho _{c}$ and $\lambda $. They correspond to either equilibrium
configurations (i.e. fulfilling eq.$\left( \ref{N1}\right) ),$ when $\lambda
=1,$ or non-equilibrium configurations, when $\lambda \neq 1.$ Now rather
than taking $\rho _{c}$ and $\lambda $ as independent variables we may use $%
N $ and $\lambda ,$ finding the binding energy as a function $E(N,\lambda ).$
It is common to change the variables to $M\equiv m_{H}N$ and $\rho
_{c}^{*}\equiv \lambda ^{-3}\rho _{c},$ so that the function should be
written $E=E(M,\rho _{c}^{*}),$ or simply $E=E(M,\rho _{c}).$ Then the
conditions of equilibrium and stability are 
\[
\partial E(M,\rho _{c})/\partial \rho _{c}=0,\;\partial ^{2}E(M,\rho
_{c})/\partial \rho _{c}^{2}>0. 
\]
But we see from these arguments that we must guarantee that $M$ is indeed
the number of baryons $N$ times a constant $m_{H}.$ The point is important
specially for the calculation of the $GR$ corrections\cite{Shapiro}.

Both eqs.$\left( \ref{57e}\right) $ and $\left( \ref{N22}\right) $ contain
three corrections to Newtonian theory. The first one derives from general
relativity and for white dwarfs is of order $GM/R_{0}\approx 10^{-4}.$ The
second is the correction for the vacuum contribution developed in this
article, which is of order $a/R_{0}^{2}\approx 10^{-5}$ (remember that $%
\left| b\right| \sim a$ $\lesssim 10$ km and $R_{o}\approx 4\times 10^{3}$
km.$)$ The third one is due to the deviation of the equation of state from
the polytropic one. The 3 corrections are small (the latter for massive
enough stars), so that we may calculate each of them as if it was alone, and
add the corrections at the end. Calculating the $GR$ correction is delicate
due to the fact that the proper volume element is not $4\pi r^{2}dr.$ This
problem however does not appear in the vacuum correction which may be
treated, in our approximation, as if we solve a purely Newtonian problem
with a density and a pressure modified by the vacuum contributions as given
in eq.$\left( \ref{7}\right) .$ I will take the corrections $GR$ and $eos$
from the literature\cite{Shapiro}, that is 
\begin{equation}
E=B\left( M_{Ch}^{2/3}-M^{2/3}\right) M\rho _{c}^{1/3}+CM\rho
_{c}^{-1/3}-DM^{7/3}\rho _{c}^{2/3},  \label{N3}
\end{equation}
where $B,C,D$ and the Chandrasekhar mass, $M_{Ch}^{2/3}$, are positive
constants (I use the notation of Ref.\cite{Shapiro}$.)$

In order to calculate the vacuum correction I begin writing the binding
energy, eq.$\left( \ref{N22}\right) ,$ taking $M^{eff}$ and $N$ from eq.$%
\left( \ref{57e}\right) $. The result should be written to first order in $G$
and $K$ (which excludes general relativistic corrections, which are of order 
$G^{2}$ or $GK$). The quantities $\rho ^{eff}$ and m$^{eff}$ are taken from
eqs.$\left( \ref{7}\right) $ and $\left( \ref{7m}\right) $ respectively,
which includes the vacuum correction to first order in the parameters $a$
and $b$. This gives 
\begin{eqnarray}
E &=&\int_{0}^{R_{o}}4\pi r^{2}dr\left[ (\rho -m_{H}n)-Gmr^{-1}m_{H}n\right]
\label{57} \\
&&+\int_{0}^{R_{o}}4\pi r^{2}dr\left[ 2a\left( \frac{2}{r}\frac{d\rho }{dr}+%
\frac{d^{2}\rho }{dr^{2}}\right) -8\pi Gar\frac{d\rho }{dr}m_{H}n\right]
+\int_{R_{o}}^{\infty }\rho _{ext}4\pi r^{2}dr,  \nonumber
\end{eqnarray}
the first term of each one of the first two integrals being the internal
energy and the second one the gravitational energy. The vacuum correction is
given by the last two integrals in eq.$\left( \ref{57}\right) .$ I shall
start estimating the third integral. The density outside the star is given
by eq.$\left( \ref{10}\right) $ with $M_{a}$ and $M_{b}$ as in eqs.$\left( 
\ref{a4}\right) $ and $\left( \ref{mb}\right) ,$ respectively. It is known
from the Lane-Endem solution of the Newtonian eq.$\left( \ref{N1}\right) $
that the density near the star surface is of the form 
\[
\rho \left( r\right) \simeq C\rho _{c}^{2}\left( \frac{G}{K}\right)
^{3/2}(R_{o}-r)^{3}, 
\]
where $C$ is a numerical constant. If this is put in the expression of $%
M_{a} $, eq.$\left( \ref{a4}\right) ,$ we get 
\[
M_{a}\simeq 24\pi C\rho _{c}^{2}\left( \frac{G}{K}\right) ^{3/2}\left(
6a+2b\right) ^{5/2}R_{o}. 
\]
Similarly we obtain 
\[
M_{b}\simeq 24\pi C\rho _{c}^{2}\left( \frac{G}{K}\right) ^{3/2}\left(
-b\right) ^{5/2}R_{o}. 
\]
If we put these expressions in the vacuum density outside the star, eq.$%
\left( \ref{10}\right) ,$ we get 
\[
\int_{R_{o}}^{\infty }\rho _{ext}4\pi r^{2}dr=8\pi C\rho _{c}^{2}\left( 
\frac{G}{K}\right) ^{3/2}\left[ \left( 6a+2b\right) ^{2}+2\left( -b\right)
^{2}\right] R_{o}^{2}. 
\]
The relevant result is that the integral is of order $a^{2},$ therefore
negliglible in comparison with the second integral of eq.$\left( \ref{57}%
\right) $ which gives therefore the main contribution to the vacuum
correction $E^{vac}$ (the first integral in eq.$\left( \ref{57}\right) $ is
the Newtonian binding energy).

The function $\rho \left( r\right) $ is the density of a $\gamma =4/3$
polytrope and, for the calculation of the integral, we may approximate $%
m_{H}n(r)\simeq \rho \left( r\right) .$ We get 
\begin{equation}
E^{vac}\simeq 2a\int_{0}^{R_{o}}4\pi rdr\left[ \frac{d^{2}\left( r\rho
\right) }{dr^{2}}-G\rho 4\pi r^{2}\frac{d\rho }{dr}\right] .  \label{58}
\end{equation}
The first term is zero (it equals $M^{vac},$ see eq.$\left( \ref{7m}\right)
) $ and the second one leads, after an integration by parts, to (see eq.$%
\left( \ref{evac}\right) $) 
\begin{equation}
E^{vac}\simeq 12\pi aG\int_{0}^{R_{o}}\rho ^{2}4\pi r^{2}dr.  \label{60s}
\end{equation}
It is interesting that the vacuum correction depends only on the parameter $%
a $, but not on $b.$

The integral eq.$\left( \ref{60s}\right) $ may be performed numerically in
terms of the central density, $\rho _{c}$ , and the Lane-Emden variables $%
\xi $ and $\theta .$ We get 
\begin{equation}
E^{vac}=12\pi aGM\rho _{c}\xi _{1}^{-2}\left| \theta ^{\prime }\left( \xi
_{1}\right) \right| ^{-1}\int_{0}^{\xi _{1}}\theta ^{6}\xi ^{2}d\xi \simeq
23.4aGM\rho _{c}.  \label{60}
\end{equation}
Adding the vacuum correction, eq.$\left( \ref{60}\right) ,$ to the standard
expression for the energy we obtain from eq.$\left( \ref{57}\right) $, 
\begin{equation}
E=B\left( M_{Ch}^{2/3}-M^{2/3}\right) M\rho _{c}^{1/3}+CM\rho
_{c}^{-1/3}-DM^{7/3}\rho _{c}^{2/3}+FM\rho _{c},  \label{61}
\end{equation}
with 
\[
F\equiv 23.4aG. 
\]
The first term in eq.$\left( \ref{61}\right) $ is the Newtonian energy, the
second is due to the departure of the equation of state from a $\gamma =4/3$
polytrope, the third one is the correction of general relativity and the
last term is the vacuum correction. For a given mass, the central density of
equilibrium is obtained when $dE/d\rho _{c}=0$, which leads to 
\begin{equation}
B\left( M_{Ch}^{2/3}-M^{2/3}\right) -C\rho _{c}^{-2/3}-2DM^{4/3}\rho
_{c}^{1/3}+3F\rho _{c}^{2/3}=0.  \label{63}
\end{equation}
Taking into account that the parameters $B,C,D$ are positive, it is easy to
see that if $F=0$ no value (positive) of $\rho _{c}$ fulfils eq.$\left( \ref
{63}\right) $ whenever $M>M_{Ch},$ that is equilibrium is not possible. In
sharp contrast a value of $\rho _{c}$ fulfilling eq.$\left( \ref{63}\right) $
exists for any mass if $F>0$. In fact the quantity in the left side of eq.$%
\left( \ref{63}\right) $ approaches +$\infty $ when $\rho _{c}\rightarrow
\infty $ and $-\infty $ when $\rho _{c}\rightarrow 0,$ so that it is zero by
continuity for some positive value of $\rho _{c}.$

For a mass, $M$, well below the Chandrasekhar limit the terms with $D$ and $%
F $ may be considered small in comparison with those with $B$ and $C,$
whence the central density of equilibrium is 
\begin{equation}
\rho _{c}\simeq \rho _{0}+\frac{3DM^{4/3}}{C}\rho _{0}^{2}-\frac{9F}{2C}\rho
_{0}^{7/3},\;\rho _{0}\equiv \left( \frac{C}{B}\right) ^{3/2}\left(
M_{Ch}^{2/3}-M^{2/3}\right) ^{-3/2}  \label{64}
\end{equation}
When $M$ approaches the Chandrasekhar limit, $M_{Ch}$, eq.$\left( \ref{64}%
\right) $ diverges and the approximations leading to it are not valid. I
shall not attempt to solve eq.$\left( \ref{63}\right) $ for $M$ close to $%
M_{Ch}$, which would be cumbersome and not very interesting. When $M>>M_{Ch}$
an approximate solution of eq.$\left( \ref{63}\right) $ is again possible,
leading to the simple expression 
\begin{equation}
\rho _{c}\simeq \left[ \frac{B\left( M^{2/3}-M_{Ch}^{2/3}\right) }{3F}%
\right] ^{3/2}\simeq 8.7\times 10^{-4}a^{-3/2}\left(
M^{2/3}-M_{Ch}^{2/3}\right) ^{3/2}.  \label{64a}
\end{equation}
However the central density, $\rho _{c}$, obtained this way is so big that
the approximations leading to it do not apply (e.g. for $M$/$M_{Ch}=1.01$ we
get $\rho _{c}\approx 10^{19}$ kg/m$^{3}$, greater than neutron star
densities.)

Stability requires that $d^{2}E/d\rho _{c}^{2}>0.$ When $dE/d\rho _{c}=0$
the stability condition leads to 
\begin{equation}
C\rho _{c}^{-1}-DM^{4/3}+3F\rho _{c}^{1/3}\geq 0.  \label{65}
\end{equation}
If $F=0,$ equilibrium is stable for any mass below a limit smaller than but
close to the Chandrasekhar mass. If $F>0$ stars may be in equilibrium for
any mass, as said above. That hydrostatic equilibrium is always stable, i.
e. for any mass, $M$, and for any value of $F>0$. In fact eq.$\left( \ref{63}%
\right) $ may be rewritten 
\[
3FM\rho _{c}^{2/3}=B\left( M^{2/3}-M_{Ch}^{2/3}\right) M+CM\rho
_{c}^{-2/3}+2DM^{7/3}\rho _{c}^{1/3}>DM^{4/3}-C\rho _{c}^{-1}, 
\]
so that eq.$\left( \ref{65}\right) $ holds true (taking into account that $%
B,C,D>0,$ $M>M_{Ch}.$) Actually the derived results for stars above the
Chandrasekhar limit are rather academic because for the probable value of $F$%
, that is when the parameter $\sqrt{a}$ lies below the limit derived in
Section 4$,$ central density of equilibrium of stars with $M$ close to or
larger than $M_{Ch}$ would be so large that the star becomes unstable
against neutronization before reaching hydrostatic equilibrium.

In summary, vacuum corrections following from the action eq.$\left( \ref{0}%
\right) $ might give rise to dramatic changes in the equilibrium and
stability of white dwarfs, but these changes would produce very small
observable effects because they are hidden by the existence of unstabilities
due to neutronization.

\section{Conclusions}

If we believe that the quantum vacuum in curved spacetime gives rise to some
stress-energy, it is plausible that the gravitational effect of this
contribution is equivalent to adding the term $aR^{2}+bR_{\mu \nu }R^{\mu
\nu }$ to the standard term, $R,$ of the Einstein-Hilbert action, at least
if the curvature scalar $R$ is not large. The presence of this term might
give rise to effects observable, in principle, in terrestrial and the solar
system observations. Then the present knowledge puts bounds on the possible
values of the parameters $a$ and $b$, namely $\sqrt{a},\sqrt{-b}<50$ km$.$
Also the vacuum stress-energy will produce some effects in the structure of
Newtonian or slightly relativistic stars. In particular in white dwarfs it
might produce stable hydrostatic equilibrium in stars above the
Chandrasekhar limit, but the effect could not be observed because such stars
would be unstable against neutronization. The effect of the vacuum
stress-energy on the internal structure of the sun would be too small to be
detected by observation. Our calculations suggest that the vacuum effects
should be important in compact bodies like neutron stars.

\textbf{Acknowledgements. }I thank an anonymous referee for valuable
criticism of a previous version of this paper. I acknowledge S. D. Odintsov
for providing useful information.

\end{document}